\documentclass[%
preprint,
 amsmath,amssymb,
 aps,
]{revtex4-2}

\usepackage{graphicx}
\usepackage{dcolumn}
\usepackage{bm}

\usepackage{graphicx,dcolumn,amssymb,amsmath,rawfonts}
\usepackage{epstopdf}
\usepackage{booktabs}
\usepackage{float}
\usepackage{xcolor}
\usepackage{theorem,latexsym}
\usepackage{natbib,enumitem}
\usepackage{multirow}
\usepackage{placeins}
\usepackage{hyperref}

\usepackage{siunitx}  
\usepackage{soul}

\usepackage{xr}
\begin{document}


\title{Charge Carrier Mobilities in $\gamma$-Graphynes: A computational approach}

\author{Elif Unsal}
\email{elif.unsal@tu-dresden.de}
\affiliation{Institute for Materials Science and Nanotechnology, TU Dresden, 01062, Dresden, Germany.}

\author{Alessandro Pecchia}
\affiliation{CNR-ISMN, Via Salaria km 29.300, 00015 Monterotondo Stazione, Rome, Italy.}

\author{Alexander Croy}
\affiliation{Institute of Physical Chemistry, Friedrich-Schiller-Universität, 07743, Jena, Germany.}


\author{Gianaurelio Cuniberti}
\email{gianaurelio.cuniberti@tu-dresden.de}
\affiliation{Institute for Materials Science and Nanotechnology, TU Dresden, 01062, Dresden, Germany.}
\affiliation{Dresden Center for Computational Materials Science (DCMS), TU Dresden, 01062, Dresden, Germany.}


\date{\today}

\begin{abstract}
Graphynes, a class of two-dimensional carbon allotropes, exhibit exceptional electronic properties, similar to graphene, but with intrinsic band gaps, making them promising for semiconducting applications. The incorporation of acetylene linkages allows for systematic modulation of their properties. However, the theoretical characterization of graphynes remains computationally demanding, particularly for electron-phonon coupling (EPC) analyses. Here, we employ the density functional tight binding method within the \textsc{DFTBephy} framework, providing an efficient and accurate approach for computing EPC and transport properties. We investigate the structural, mechanical, electronic, and transport properties of graphynes, comparing transport calculations using the constant relaxation-time approximation and the self-energy relaxation-time approximation (SERTA) alongside analytical models based on parabolic- and Kane-band approximations. For graphyne, the SERTA relaxation time is 0.63 (1.69) ps for holes (electrons). In graphdiyne, the relaxation time is 0.04 (0.14) ps for holes (electrons). While the hole mobilities in graphyne are on the order of 10$^3$ cm$^2/$Vs, the electron mobilities reach up to 10$^4$ cm$^2/$Vs. In graphdiyne, the mobility values for both types of charge carriers are on the order of 10$^2$ cm$^2/$Vs. The phonon-limited mobilities at room temperature in graphyne fall between those of graphene and MoS$_2$, while in graphdiyne, they are comparable to those of MoS$_2$. 
\end{abstract}

\maketitle


\section{Introduction}
\label{sec:intro}

Graphynes offer promising electronic properties and are potentially interesting materials for many semiconducting applications~\cite{Diederich-2010, Hirsch-2010}. Unlike graphene, which is gapless and thus limited in particular electronic and optoelectronic devices~\cite{graphene-device-limitations-2010}, graphynes exhibit tunable electronic band-structures with intrinsic band-gaps, facilitated by the hybridization of sp-, sp$^2$-, and sp$^3$-bonded carbon atoms. These unique hybridized bonds allow for the creation of extended 2D planar networks that share the same P6/mmm symmetry as graphene~\cite{g-GY-Eckhardt-prediction-87}. While graphynes exhibit lower values for cohesive energy, planar packing density, in-plane stiffness, and Fermi velocities compared to graphene, they offer advantages like a higher Poisson's ratio and a wider range of pore sizes~\cite{g-GY-theo-emass}. While $\alpha$-, $\beta$-, and $6,6,12$-graphynes demonstrate higher electronic transmittance compared to graphene, $\gamma-$graphyne achieves the highest thermoelectric efficiency nearly an order of magnitude greater than that of graphene.\cite{Sevincli2014-gy}

\begin{figure}[t!]
\centering
\includegraphics[width=16.0cm]{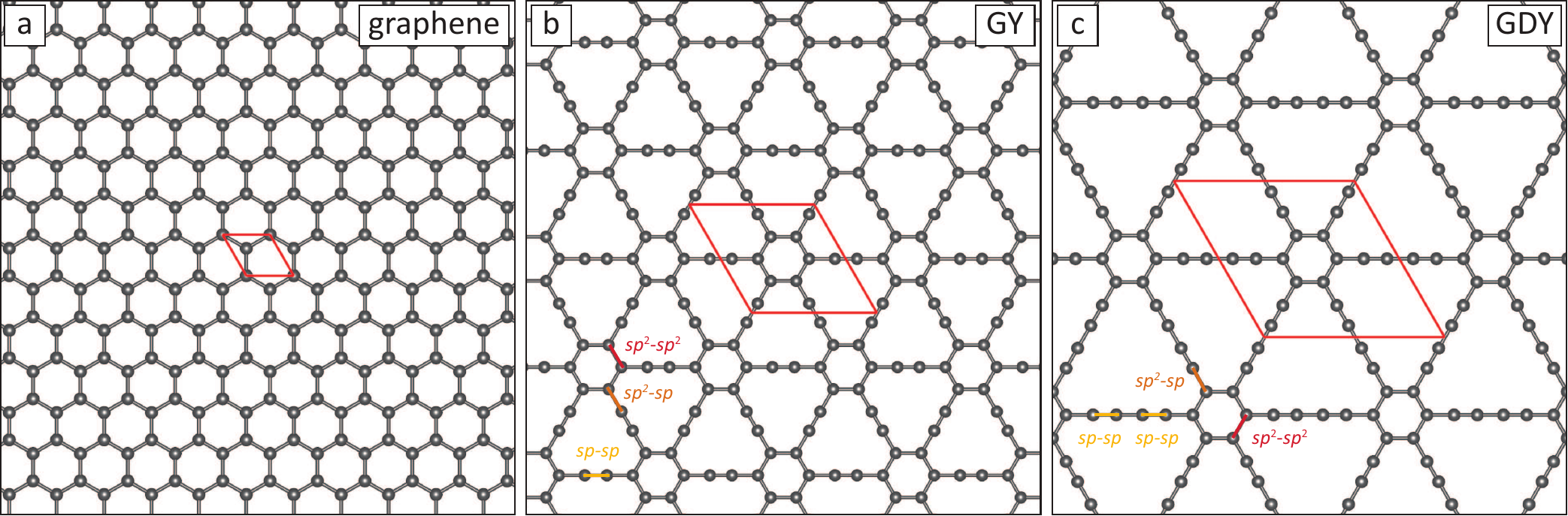}
\caption{Optimized geometries the monolayers of (a) graphene, (b) graphyne (GY) and (c) graphdiyne (GDY). The red rhombi represent the unit cells.}
\label{FIG:structures}
\end{figure}

Incorporating acetylene linkages in graphynes provides a versatile approach to tailoring their properties. The adjustable length of such acetylene chains systematically modifies the structural arrangement and influences the material's electronic, mechanical, and transport properties~\cite{Diederich-2010,g-GY-Eckhardt-prediction-87}. This tunability parallels the customizable architectures of covalent organic frameworks (COFs)~\cite{COFs-Yaghi,COFs-Colson} and 2D Fused Aromatic Networks (FANs)~\cite{FANs}, where rigid aromatic backbones enable precise control over porosity and functionality.
Such structural versatility makes these materials promising for applications also in catalysis and energy storage~\cite{FANs,Ivanovskii2013-gasstorage,intro-sensing2023}.
The uniform distribution of nanoscale pores also makes these materials highly suitable for nanoelectronics, including molecular sieving for gas separation\cite{GY-sensing}, water desalination\cite{GY-water-desalination}, and biomedical usage such as cholesterol extraction\cite{GY-cholesterol}.

The theoretical characterization of these materials presents a significant challenge, primarily due to their large unit cells. This complexity demands substantial computational resources, especially when analyzing electron-phonon couplings (EPCs), which are essential for understanding their electronic transport properties\cite{Ponce2020,Lundstrom}.
Using semi-empirical methods such as density functional tight binding (DFTB)~\cite{DFTB} instead of density functional theory (DFT) offers a practical solution. DFTB effectively reduces the computational demands of DFT, providing a reasonable balance between efficiency and accuracy~\cite{defected-GR-intro,COF-intro}. Here, we employed the \textsc{DFTBephy} Python package, which is developed to compute EPCs within a DFTB framework~\cite{DFTBephy}. It enables the calculation of transport properties within the Boltzmann transport framework in the relaxation-time approximation, integrating these calculations into an efficient computational workflow. \textsc{DFTBephy} offers a promising approach for addressing the computational challenges associated with large and complex systems, enhancing the accessibility of EPCs and transport property analyses.

In addition to computational approaches, analytical models provide valuable insights into charge carrier transport. While the constant relaxation-time approximation (CRTA) is often used for its simplicity, it does not always capture non-parabolic effects that become significant at higher carrier concentrations~\cite{Lundstrom}. The Kane-band model offers a more accurate representation of charge carrier mobility in such cases~\cite{Walsh2019}. Although these analytical models may not always achieve the accuracy of computational methods, they offer a fast and efficient way to estimate mobilities and conductivities in complex materials. This makes them particularly useful for preliminary studies and identifying trends before engaging in more demanding simulations. These analytical approaches provide a comprehensive framework for understanding charge transport in complex materials.


\section{Methodology}
\label{sec:method}

\subsection{Electron-Phonon Couplings}\label{subsec:EPC}

\begin{figure}
\centering
\includegraphics[width=16.0cm]{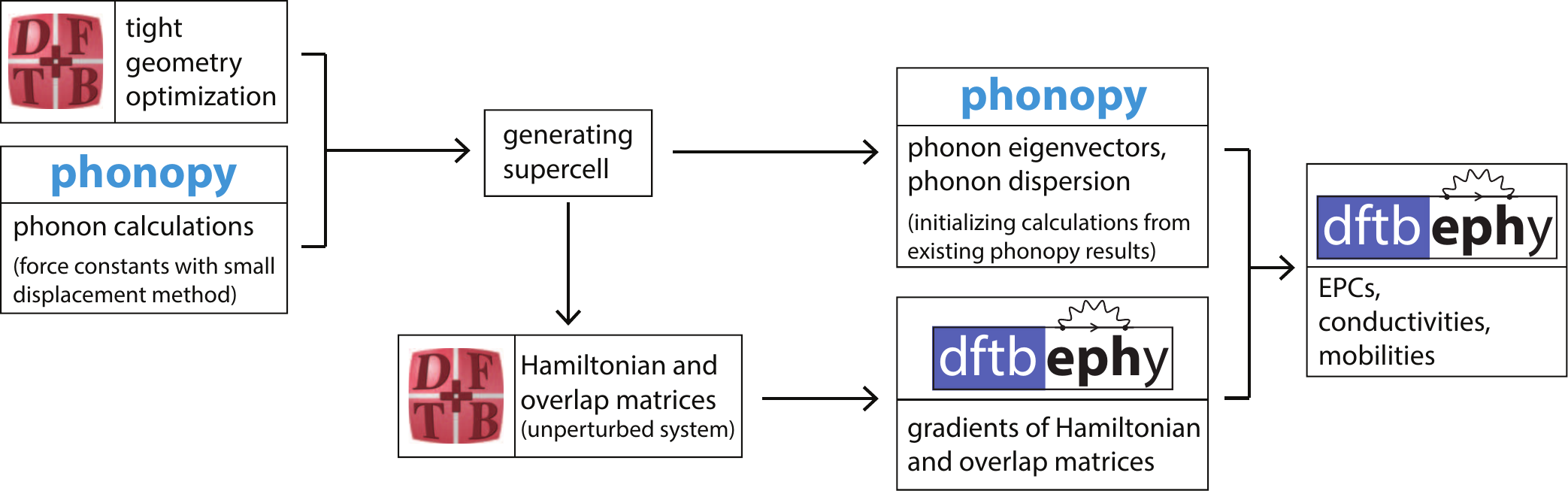}
\caption{ Illustration of the workflow for calculating electron-phonon couplings and transport properties using the \textsc{DFTBephy} code.}
\label{fig:workflow}
\end{figure}

This section briefly describes the calculations of electron-phonon couplings with \textsc{DFTBephy}~\cite{DFTBephy}. The workflow is summarized in Fig.\ \ref{fig:workflow}. Before starting EPC calculations, it is essential to perform tight geometry optimization and phonon calculations. 

The calculation of the EPC in \textsc{DFTBephy} involves multiple steps. The first step is to start with creating a super-cell, which is analogous to the way utilized in \textsc{Phonopy}~\cite{phonopy-Togo2015,phonopy-Togo_2023}. \textsc{DFTBephy} uses the \textsc{DFTB+}~\cite{DFTB+} code to obtain the real-space Hamiltonian and Overlap matrices, $H_{\mu,\nu}(\ell s, \ell' s')$ and $S_{\mu,\nu}(\ell s, \ell' s')$, where \textit{l} denotes the cell index, and \textit{s} is the sublattice index of each atom. These matrices are then Fourier-transformed with respect to the cell indices, yielding $H_{\mu s,\nu s'}(\vec{k})$ and $S_{\mu s,\nu s'}(\vec{k})$.

The electronic properties are necessary for EPC calculations. After solving a generalized eigenvalue problem for each \textit{k}-point, the electronic band structure $\varepsilon_n(\vec{k})$ and the corresponding eigenstates $U_{\nu s',m}(\vec{k})$ are obtained,
\begin{equation}\label{eq:band-structure}
    \varepsilon_{n}(\vec{k}) = \sum_{\tau,\mu}\sum_{s,s''} U^*_{\mu s, n}(\vec{k}) H_{\mu s,\nu s'}(\vec{k}) U_{\nu s', n}(\vec{k}) \;.
\end{equation}
If required, the band velocities $\vec{v}_n(\vec{k})$ can also be calculated using the following equation,
\begin{equation}\label{eq:velocity}
    \vec{v}_n(\vec{k}) = \frac{1}{\hbar}\frac{\partial \varepsilon_{n}(\vec{k})}{\partial \vec{k}}
    = \frac{1}{\hbar}\sum_{\tau,\mu}\sum_{s,s''} U^*_{\mu s, n}(\vec{k}) \left[\frac{\partial H_{\mu s,\nu s'}(\vec{k})}{\partial \vec{k}}
    - \varepsilon_{n}(\vec{k})\frac{\partial S_{\mu s,\nu s'}(\vec{k})}{\partial \vec{k}}\right]U_{\nu s', n}(\vec{k})\;.
\end{equation}

In the second step, each atom in the super-cell is displaced. Then, in a finite-difference scheme, the real-space gradients of the Hamiltonian and overlap matrices are evaluated at the unperturbed positions.
The first-order change of $H_{\mu,\nu}(\ell s, \ell' s')$ and $S_{\mu,\nu}(\ell s, \ell' s')$ is calculated by noting that the two-center matrix elements depending on the distance vector, not individual positions. Then the gradients are also Fourier transformed yielding ${\partial H_{\mu s,\nu s'}(\vec{k})}/{\partial \vec{R}}$ and ${\partial S_{\mu s,\nu s'}(\vec{k})}/{\partial \vec{R}}$.

EPC calculations also require phonon frequencies and eigenvectors. The phonon dispersion is determined from the dynamical matrix $D_{\alpha \beta}(ss',\vec{q})$~\cite{phonopy-Togo2015,phonopy-Togo_2023},
\begin{equation}
    D_{\alpha \beta}(ss',\vec{q}) = \frac{1}{\sqrt{m_s m_{s'}}} \sum_{l'} \Phi_{\alpha \beta} (s0,s'l')
    e^{\imath \vec{q}\cdot [\vec{R}_{s'l'}-\vec{R}_{s0}]}\;, 
\end{equation}
where $m_s$ is the atomic mass of atom $s$, $\vec{q}$ is the phonon wave vector, $\alpha$ and $\beta$ denote the Cartesian coordinates, and $\Phi_{\alpha \beta} (s0,s'l')$ are the harmonic force constants. The phonon frequency $w_{\lambda}(\vec{q})$ and eigenvectors $\vec{\xi}^\lambda_s(\vec{q})$ are determined by solving the eigenvalue equation of the dynamical matrix,
\begin{equation}
    \sum_{s'\beta} D_{\alpha \beta}(ss',\vec{q}) \vec{\xi}^\lambda_{\beta s'} (\vec{q})= [\omega_{\lambda}(\vec{q})]^2 \vec{\xi}^\lambda_{\alpha s} (\vec{q})\;,
\end{equation}
where $\vec{\xi}^\lambda_s(\vec{q})$ is the eigenvector of atom $s$ in phonon branch $\lambda$, and $w_{\lambda}(\vec{q})$ is the corresponding phonon frequency.

Finally, the EPC matrix is calculated from the expression
\begin{multline}\label{eq:EPCs}
 g^{\lambda}_{nm}(\vec{k},\vec{q}) = \sqrt{\frac{\hbar}{2 \omega_{\lambda}(\vec{q})} }  \sum_{s,\mu,s',\nu} U^{*}_{n,s\mu}(\vec{k}+\vec{q})  
 \times \left\{ \left[ \frac{\partial H_{s \mu, s' \nu}}{\partial \vec{R}_s} (\vec{k}) \right.
 \left.- \varepsilon_m(\vec{k}) \frac{\partial S_{s \mu, s' \nu}}{\partial \vec{R}_s}(\vec{k})\right]
  \frac{\vec{\xi}^{\lambda}_s(\vec{q})}{\sqrt{m_s}} \right. \\
  - \left. \left[ \frac{\partial H_{s \mu, s' \nu}}{\partial \vec{R}_s} (\vec{k}+\vec{q}) \right.
 \left.- \varepsilon_n(\vec{k}+\vec{q}) \frac{\partial S_{s \mu, s' \nu}}{\partial \vec{R}_s}(\vec{k}+\vec{q})\right]  \frac{\vec{\xi}^{\lambda}_{s'}(\vec{q})} {\sqrt{m_{s'}}}  \right\} 
  U_{s'\nu,m}(\vec{k})\;.
\end{multline}
In addition to the electronic structure and the phonons, the calculation of EPCs requires the computation of the (real-space) gradients of the Hamilton and the overlap matrix. Those can be obtained via a finite displacement method.
For more detailed information about the derivation and the calculation scheme, please see Ref.~\cite{DFTBephy}.


\subsection{Relaxation Times}\label{subsec:Relaxation-Times}
To obtain the relaxation time from the EPC constants, we use the SERTA scattering rate~\cite{Ponce2020},
\begin{equation}\label{eq:tau_SERTA}
    \tau^{-1}_{n}(\vec{k}) = \sum_m \int \frac{d^3 q}{\Omega_{BZ}}  \tau^{-1}_{nm}(\vec{k},\vec{k}+\vec{q})\;.
\end{equation}
The partial decay rate $\tau^{-1}_{nm}(\vec{k},\vec{k}+\vec{q})$ is given in terms of the electron-phonon coupling matrix $g^\lambda_{mn}(\vec{k}, \vec{q})$ and the occupation factors of the involved states,
\begingroup\makeatletter\def\f@size{10.5}\check@mathfonts
\begin{align}
    \tau^{-1}_{nm}(\vec{k},\vec{k}+\vec{q}) ={}& \frac{2\pi}{\hbar} \sum_{\lambda} |g^\lambda_{mn}(\vec{k}, \vec{q})|^2 \left[\left(n(\omega_{\lambda}(\vec{q})) + 1 - f^0(\varepsilon_m(\vec{k}+\vec{q}))\right) \delta(\varepsilon_m(\vec{k}+\vec{q})-\varepsilon_n(\vec{k})+\hbar\omega_{\lambda}(\vec{q})) \right. \notag\\
    {}&\left.+
    \left(n(\omega_{\lambda}(\vec{q})) + f^0(\varepsilon_m(\vec{k}+\vec{q}))\right) \delta(\varepsilon_m(\vec{k}+\vec{q})-\varepsilon_n(\vec{k})-\hbar\omega_{\lambda}(\vec{q}))\right]\;.
\end{align}
\endgroup
Here, $f^0(\varepsilon)$ and $n(\omega)$ denote the Fermi function and the Bose-Einstein distribution, respectively, which characterize the occupation of electronic states and the phonons in equilibrium.
Within \textsc{DFTBephy} the partial decay rates are computed using a Gaussian smearing function with constant width instead of the $\delta$-functions. The integral over the Brillouin zone in Eq.\ \eqref{eq:tau_SERTA} is replaced by a sum over q-points.


\subsection{Charge Transport Properties}\label{subsec:Transport}

Here, we investigate the transport properties of charge carriers in the semi-classical picture using the Boltzmann transport equation~\cite{Ponce2020}. The conductivity tensor is defined as 
\begin{equation}\label{Eq:sigma}
    \sigma_{\alpha\beta} = \frac{e^2}{A_{uc}} \sum_n \int \frac{d^2k}{\Omega_{BZ}} \left[-\frac{\partial f^0(\varepsilon_{n}(\vec{k}); \mu, T)}{\partial \varepsilon}\right]  v_{n\alpha}(\vec{k}) v_{n\beta}(\vec{k}) \tau_{n}(\vec{k})\;,
\end{equation}
where $A_\text{uc}$ is the area of the unit cell and $v_{n\alpha}(\vec{k})$ and $v_{n\beta}(\vec{k})$ are the charge carrier group velocities in the directions $\alpha$ and $\beta$, which can be calculated via Eq.\ \eqref{eq:velocity}. The charge carrier densities $n_\text{c}$ at a given temperature $T$ and a chemical potential $\mu$ can be defined as
\begin{equation}\label{Eq:N_c}
    n_c = \frac{1}{A_{uc}}\sum_n \int \frac{d^2k}{\Omega_{BZ}}[f^0(\varepsilon_n; \mu,T) -f^0(\varepsilon_n;\varepsilon_F,0)]\;,
\end{equation}
where the electron density, $n_c=n_{el}$, is calculated by summing over the conduction bands, while for the holes, $n_c=n_{h}$, the sum is taken over the valence bands.


\subsection{Charge Transport within the Non-parabolic Dispersion Framework}\label{subsec:Non-parabolicity}

The conduction and valence bands of semiconductors are typically assumed to be parabolic around the band edges. However, the band structure of many materials deviates from this simple parabolic model in practice, especially considering the large carrier concentrations~\cite{Lundstrom,Walsh2019}. The Kane model is a widely used method for modelling non-parabolicity in electronic dispersion~\cite{Maassen} and can be defined via
\begin{equation}
    \varepsilon + \alpha\varepsilon^2 = \frac{\hbar^2 \vec{k}^2}{2m^*}\;, 
\end{equation}
where $m^*$ is the effective mass and $\alpha$ is the non-parabolicity parameter. The corresponding band velocity along direction $a$ is defined as
\begin{align}\label{Eq:velocity-Kane}
    v_{a} =& \frac{1}{\hbar} \frac{\partial \varepsilon}{\partial k_{a}} = \frac{\hbar k_a}{m^*}\left( \frac{1}{2\alpha\varepsilon +1}\right)\;.
\end{align}

In the Kane model, the bands are flattened, and the density of states increases by a factor of $(1+2\alpha\varepsilon)$ compared to the parabolic approximation. For the charge carrier densities, we obtain
\begin{align}\label{Eq:N_C-Kane}
    n_c= \frac{m^* g_v}{2\pi \hbar^2} \int\  d\varepsilon (1+2\alpha\varepsilon)[f^0(\varepsilon; \mu,T) -f^0(\varepsilon;\varepsilon_F,0)]  \;,
\end{align}
where $g_v$ is the valley degeneracy. The conductivity becomes 
\begin{equation}\label{Eq:sigma-Kane}
\sigma_{2D} = \frac{e^2 g_v}{2 \pi \hbar^2} \int d\varepsilon \left[-\frac{\partial f^0(\varepsilon(\vec{k}); \mu, T)}{\partial \varepsilon}\right] \frac{\varepsilon + \alpha \varepsilon^2}{1 + 2\alpha\varepsilon } \tau_{n}(\vec{k})\;,  
\end{equation}
where $\tau$ is the $k$-dependent relaxation time. The alpha value differs for each band. Here, we consider a single band. If multiple bands are involved, their contributions need to be summed up. The contributions of valley degeneracies must be taken into account, depending on the high-symmetry point at which the band edge is located. We averaged the trace of the conductivity tensor considering only the in-plane conductivities, $(\sigma_{xx}+\sigma_{yy})/2$. Details of the derivation can be found in the Supporting Information (SI).


\subsection{Computational Details}\label{subsec:compt-details}

We used the \textsc{mio-1-1}~\cite{mio-SCC} parametrization set without self-consistent charges for structure optimization with DFTB. The maximum force tolerance was set to $10^{-6}$ a.u., and a temperature of 100 K was applied for Fermi filling. For GY and GDY, $21 \times 21 \times 1$ and $15 \times 15 \times 1$ $k$-meshes were used, respectively. Details regarding the comparisons of different DFTB parametrization sets can be found in the SI file.
In all calculations, 15 \AA{} vacuum was inserted to avoid interactions between the adjacent layers. 

In parallel, we performed DFT calculations by using the Vienna ab-initio Simulation Package (VASP)~\cite{vasp1,vasp2} for geometry optimization and band structure calculations. The Perdew-Burke-Ernzerhof (PBE)~\cite{PBE} form of GGA was adopted to describe electron exchange and correlation. The energy cut-off for plane-wave expansion was set to 525 eV. The Gaussian smearing method was employed for the total energy calculations and the width of the smearing was chosen as 0.05 eV. The energy difference between successive electronic steps was set to 10$^{-8}$ eV. For the primitive unit cells of GY and GDY, $24 \times 24 \times 1$ and $21 \times 21 \times 1$ $\Gamma$-centered $k$-point samplings were used, respectively. 

Vibrational properties are investigated by using the small displacement method as implemented in {\scshape{Phonopy}} code~\cite{phonopy-Togo2015,phonopy-Togo_2023}. To obtain force constants, 7$\times$7$\times$1 and 5$\times$5$\times$1 supercells were generated for graphyne (GY) and graphdiyne (GDY), respectively. Atoms were displaced by 0.001 \AA. Tolerance to find the symmetry of crystal structure was set to 10$^{-4}$ for DFTB calculations and 10$^{-5}$ for DFT calculations. 

We calculated the scattering rates within the SERTA (see Sec.\ \ref{subsec:Relaxation-Times}) by setting the chemical potentials to the band edges at 300 K. A $16 \times 16 \times 1$ $k$-mesh and a $100 \times 100 \times 1$ $q$-mesh were employed for the calculations. For the linear interpolation of the scattering rates (for the analytical Kane-mobilities), we used the \textit{interp1d} function from the \textsc{SciPy} library.

For the conductivity calculations, a $150 \times 150 \times 1$ $k$-mesh and a $100 \times 100 \times 1$ $q$-mesh were used. For GY, the $k$-vector was shifted at the $M-$point (band edge) and only two bands were considered. For GDY, the $k$-vector was located at the $\Gamma-$point (band edge). In this case, four bands were considered. In the calculations for both materials, a chemical potential range of 0.3 eV was employed, starting from slightly below (above) the conduction (valence)  band-edge and extending into the respective bands.


\section{Results and Discussion}
\label{sec:results-and-discussion}

\begin{figure}[t]
\centering
\includegraphics[width=16.0cm]{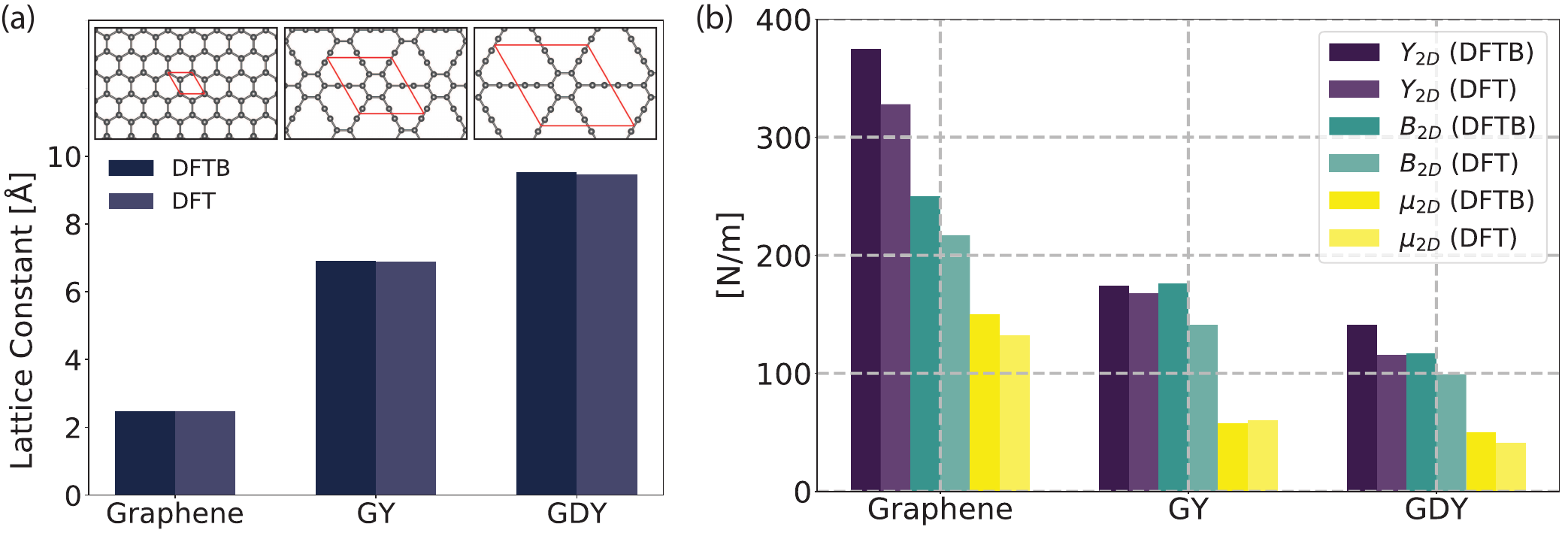}
\caption{ (a) Comparison between lattice constants calculated with DFTB and DFT for graphene, graphyne (GY) and graphdiyne (GDY). Geometric structures are shown as inset figure. Red rhombuses represent the unit cells. (b) Comparison between the 2D Young’s modulus $Y_{2D}$, bulk modulus $B_{2D}$, and shear modulus $\mu_{2D}$ calculated with DFTB and DFT.} 
\label{FIG:DFTB-DFT-comparison}
\end{figure}

\begin{figure}
\centering
\includegraphics[width=16.0cm]{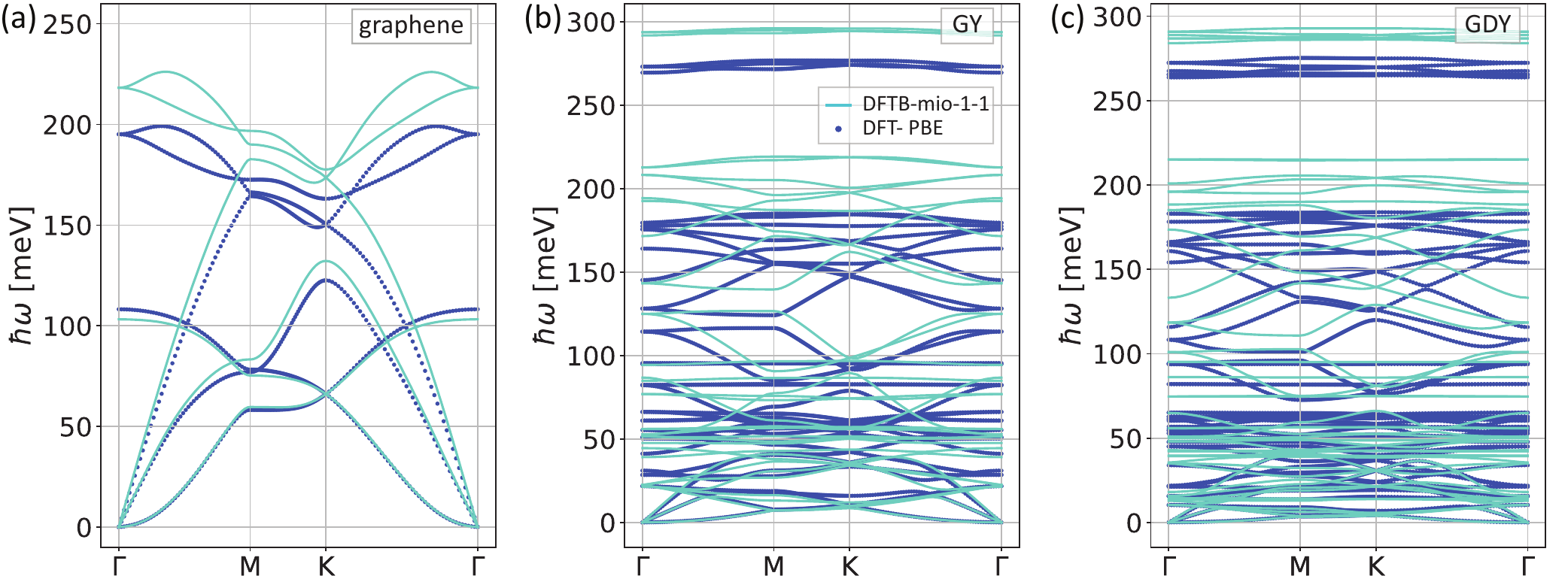}
\caption{ The phonon band spectra for the monolayers of (a) graphene, (b) graphyne (GY) and (c) graphdiyne (GDY) along the high symmetry points $\Gamma$-$M$-$K$-$\Gamma$. Supercells of $7\times 7 \times 1$ were constructed for graphene and GY, while a $5 \times 5 \times 1$ supercell was used for GDY.}
\label{FIG:phonon-bands}
\end{figure}

\begin{figure}
\centering
\includegraphics[width=16.0cm]{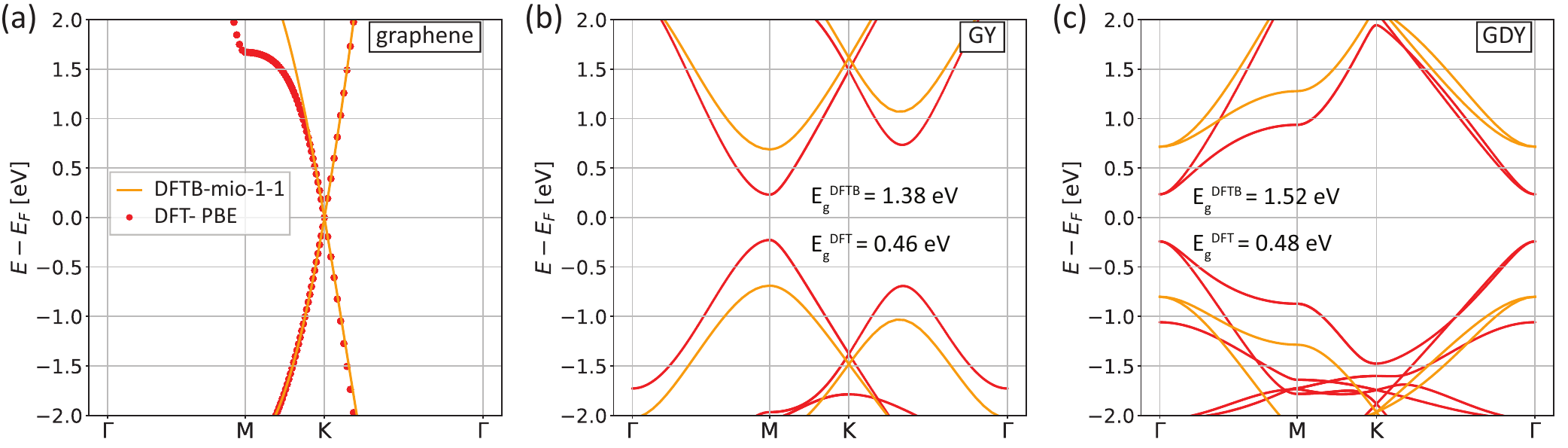}
\caption{ The electronic band structures of (a) graphene, (b) graphyne (GY), and (c) graphdiyne (GDY). }
\label{fig:2_electronic_bands}
\end{figure}

\paragraph*{Structural properties}Monolayer $\gamma$-graphynes exhibit a planar structure and are categorized under the $P6/mmm$ point group symmetry. The unit cell of GY contains 12 carbon atoms, while GDY's unit cell contains 18 atoms. Fig. \ref{FIG:structures} shows the DFTB-optimized geometries of graphene, GY and GDY. The lattice constants are $2.47$, $6.92$ and $9.53$ {\AA}, respectively. A comparison with computed literature values, given in Fig.\ \ref{FIG:DFTB-DFT-comparison}(a) and in Table \ref{table-SI-structural-prop.} in the SI file, shows a very good agreement for lattice constants and bond lengths~\cite{g-GY-theo-emass,g-GY_theo-bonds}.
\paragraph*{Phonons and elastic properties}Based on the equilibrium structures, the phonon dispersions were calculated (see Sec.\ \ref{sec:method}). The analysis reveals that the calculated GY and GDY structures are dynamically stable (see Fig.\ \ref{FIG:phonon-bands}). The DFTB results are in good agreement with DFT up to $\approx 23$ meV. The maximum frequencies of the optical phonon branches at $\Gamma$ is around $290$ meV, corresponding to the stretching mode present in all the structures. In contrast to graphene, these modes in graphyne structures are not dispersive. 
From the acoustic branches of the phonon dispersions, the stiffness tensor elements $C_{11}$ and $C_{12}$ can be extracted. Due to the honeycomb lattice structure, they are related to the sound velocities of the LA and TA modes via $C_{11}= \rho v_{\rm LA}^2$ and $C_{12}= \rho v_{\rm TA}^2$, where $\rho$ is the mass density and the sound velocities are given by the slope of the respective branch at $\Gamma$. The in-plane elastic properties, 2D Young's modulus $Y_{2D}$, bulk modulus $B_{2D}$ and shear modulus $\mu_{2D}$, of the structures were calculated from these stiffness constants as in Ref.~\cite{ACroy2022}. The agreement of the DFTB results with DFT calculations is again quite good, as seen in Fig.\ \ref{FIG:DFTB-DFT-comparison}(b) (and the calculated values can be found in Table~\ref{table-SI-structural-prop.} in the SI file). The in-plane stiffness of graphyne sheets decreases rapidly due to the introduction of acetylene linkers. Although triple bonds (sp) in acetylene are stronger than aromatic bonds (sp$^2$), their association with weaker single bonds (sp$^3$) leads to a reduction in strength as the number of increasing acetylene linkages~\cite{Buehler2012}. In this context, we observe a similar trend, with the in-plane stiffness of GDY being lower than that of graphene. This trend is also observed for bulk modulus $B_{2D}$, and shear modulus $\mu_{2D}$.
\paragraph*{Electronic structure}A comparison of the electronic band-structure calculations with DFTB and DFT is shown in Fig.\ \ref{fig:2_electronic_bands}. While graphene is a semi-metal, GY and GDY are direct band-gap semiconductors. For GY, the band edges are located at the $M$-point, resulting in a valley degeneracy of three ($g_v= 3$), and the Dirac nodal lines are around $E_{\rm F}-1.49$ and $E_{\rm F} + 1.61$ eV at the $K$-point. The band edges of GDY are located at the $\Gamma-$point, with degeneracy observed in both the conduction and valence band edges. As the band edge lies at the $\Gamma-$point, valley degeneracy is absent.
The band-gap value of GY calculated with DFTB is $1.38$ eV, which is comparable to the value predicted within the extended Hückel theory and with an empirical correction which is $1.2$ eV~\cite{g-GY-Eckhardt-prediction-87}. The gap values obtained with these methods are $\approx 0.9$ eV higher than those obtained with bare DFT. This discrepancy likely arises from using a GGA-PBE functional, known to underestimate band gaps, particularly in systems with delocalized electrons~\cite{Perdew2017}. A similar trend is observed for GDY, where PBE predicts a band gap of approximately $0.5$ eV, while DFTB predicts $\approx 1.5$ eV. The band gap values obtained by DFT for GY and GDY are in good agreement with the literature~\cite{g-GY-theo-emass,Gao2011-GDY-mu}.

For the effective mass calculation for DFT bands, we used \textsc{effmass} Python package~\cite{effmass-Whalley2018}. 
In GY, DFT calculations indicate that the effective masses of electrons and holes are smaller in the $M-K$ direction ($m_e=0.08$ $m_0$, $m_h=0.09$ $m_0$) compared to the $M-\Gamma$ direction ($m_e=0.20$ $m_0$, $m_h=0.22$ $m_0$). Our results align well with the results reported in Ref.\ ~\cite{g-GY-theo-emass}.
On the other hand, in GDY, the effective masses do not vary with direction ($m_e = m_h = 0.09$ $m_0$ for the primary bands, while $m_e = m_h = 0.07$ $m_0$ for the secondary bands).
In both materials, the effective masses from DFTB bands show minimal directional variation, a consequence of the isotropic nature of the band structure in both directions. In GY, the effective masses from the DFTB band structure exhibit slight anisotropy between the $M-K$ ($m_e = 0.24$ $m_0$ and $m_h = 0.26$ $m_0$) and $M-\Gamma$ direction ($m_e = 0.30$ $m_0$ and $m_h = 0.32$ $m_0$). In GDY, the primary bands have isotropic effective masses in both directions: $m_e = 0.28$ $m_0$ and $m_h = 0.32$ $m_0$. For the secondary bands, the effective masses can also be assumed to be isotropic: $m_e = 0.19$ $m_0$ and $m_h = 0.21$ $m_0$ in the $M-K$ direction and $m_e = 0.18$ $m_0$ and $m_h = 0.20$ $m_0$ in the $M-\Gamma$ direction. Although the effective masses obtained from DFTB are larger than those from DFT, both methods consistently indicate isotropic behaviour with respect to directional dependence.
The effective mass values are also given in Table~\ref{table-SI-electronic-prop.} in the SI file.

\begin{figure}
\centering
\includegraphics[width=16.0cm]{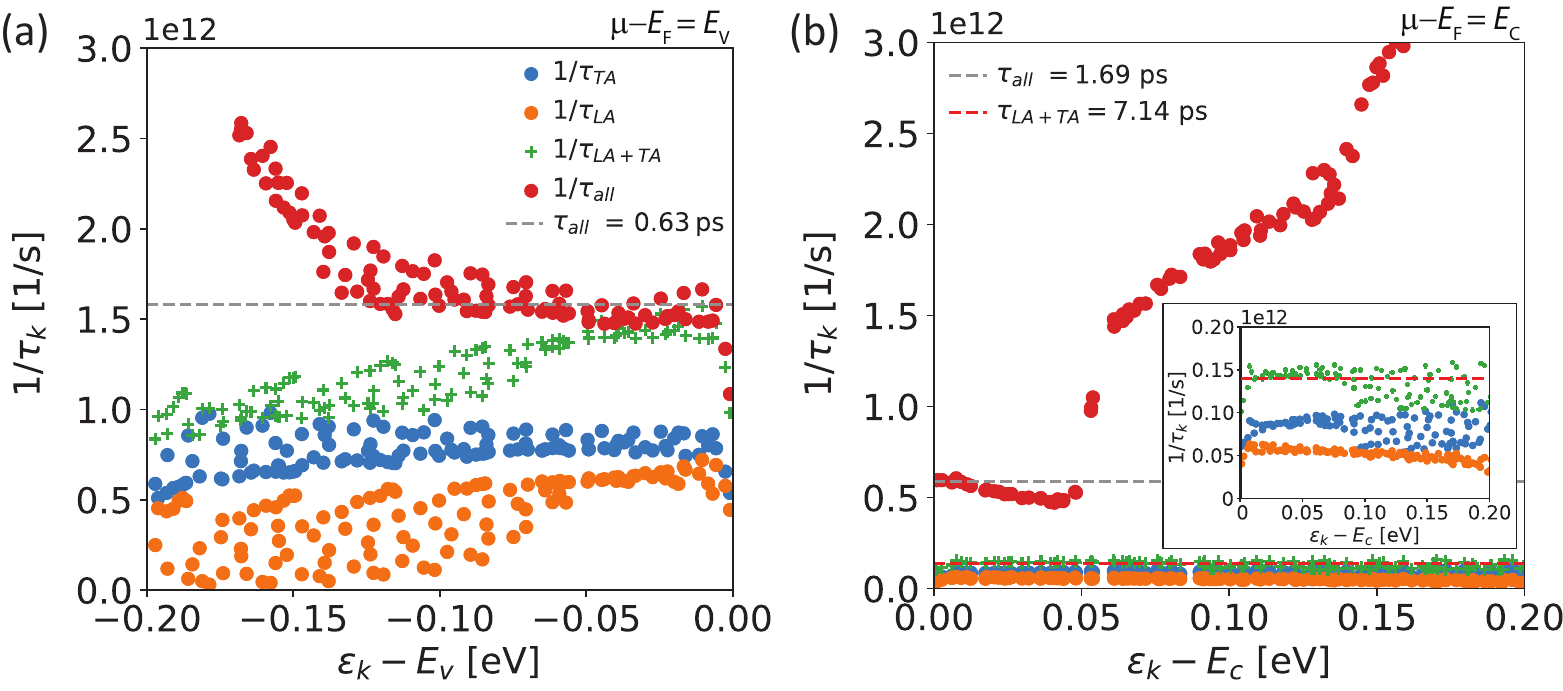}
\caption{ The scattering rates (inverse life-times) in this plot calculated at fixed temperature $T = 300$ K and chemical potential within SERTA for (a) the holes and (b) the electrons in monolayer graphyne (GY). The chemical potentials were set to the band edge values. }
\label{fig:g-graphyne-relaxation-times}
\end{figure}

\paragraph*{Scattering rates}Scattering processes were considered exclusively from the conduction and valence bands, with two bands included for GY and four bands for GDY.
In Fig.\ \ref{fig:g-graphyne-relaxation-times}, the scattering rates for electrons and holes in GY are shown as a function of energy. When all scattering is included, the relaxation times for electrons and holes are $1.69$ and $0.63$ ps, respectively (see Table \ref{table:relaxation-times}). The calculated values for electrons and holes are 54~\% and 21~\% higher than those in the literature, respectively~\cite{JXi2014}. The contribution to the scattering around the band edge for the holes is primarily from acoustic phonon scattering (see Fig.\ \ref{fig:g-graphyne-relaxation-times}(a)). The scatterings from acoustic phonons remain constant over an energy range of $0.2$ eV. However, moving further into the valence band, we see that scattering from optical phonons becomes significant. As for electrons, the contribution from optical phonons is roughly three times larger than that from acoustic phonons. Similar to the holes, acoustic phonon scattering remains relatively constant within the conduction band. The scatterings from optical phonons increase significantly as we move deeper into the band. We compared the charge carrier scatterings in GY and GDY in Fig \ref{fig:SI_scatterings} in the SI file.
In GDY, degeneracy is present in both valence and conduction bands at the $\Gamma$-point and the scattering rates were calculated by considering these degenerate bands.  The relaxation time values provided in Table \ref{table:relaxation-times} were calculated using Matthiessen’s Rule, which combines the individual contributions from each band, expressed as $1/\tau = 1/\tau_{band1} + 1/\tau_{band2}$.
When the chemical potential is at the band edge, the total scattering for the electrons is approximately 2.5 times larger than that of holes. Acoustic phonon scattering rates in GDY display a similar trend, remaining constant within the bands for both electrons and holes, whereas optical phonon rates show a notable energy dependency. The total scattering rates in GDY are one order of magnitude larger than those in GY.
\begin{figure}
\centering
\includegraphics[width=16.0cm]{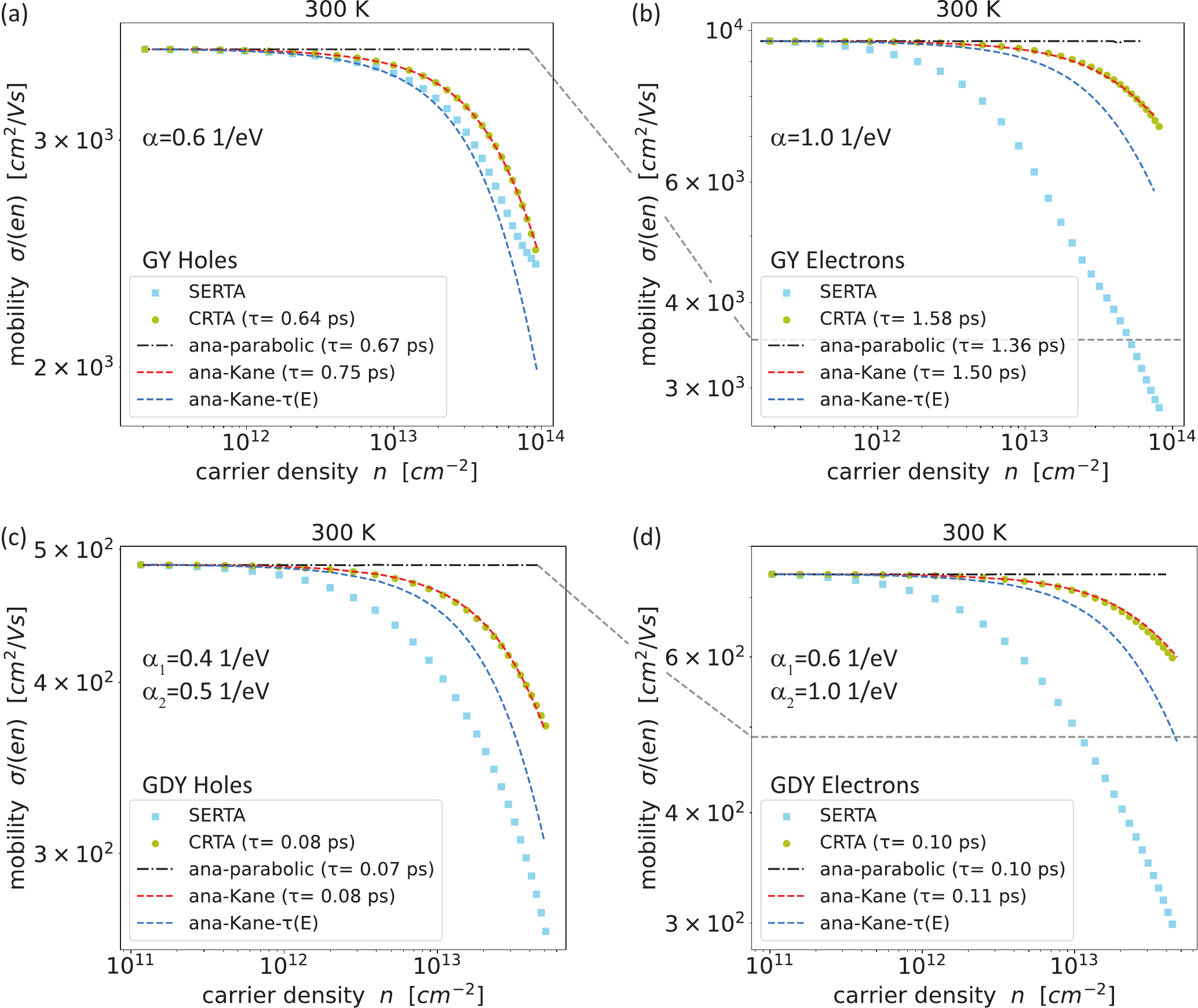}
\caption{Mobilities as a function of carrier concentrations for graphyne (GY): (a) holes and (b) electrons. Mobilities of (c) holes and (d) electrons in graphdiyne (GDY). The mobilities were scaled based on values obtained using the SERTA.}
\label{fig:g-graphyne-mobilities}
\end{figure}
\paragraph*{Phonon-limited mobility}Next, we calculated the room-temperature mobilities using SERTA and CRTA. For the latter, the relaxation times are found by matching the low-density mobilities of SERTA. Additionally, we calculate analytical CRTA mobilities by considering single parabolic-band and Kane-band~\cite{Lundstrom} models. For the latter, we obtained the non-parabolicity parameter $\alpha$ by fitting the energy dispersion close to the band edges. In both materials, $\alpha$ for electrons is larger than for holes, indicating that the effective mass of electrons depends strongly on energy. The results with SERTA provide the most detailed analysis because they factor in both the actual band structure and the calculated scattering rates.
For deformation potential calculations in 2D, it is generally assumed that the bands are parabolic and that the relaxation time remains constant due to the characteristics of the density of states. In this case, the mobility remains constant, regardless of the carrier density. The band structure and relaxation times play crucial roles in determining mobility, so it is important to calculate them accurately, as any variation in mobility can be attributed to changes in both. Parabolic-band and Kane-band models within CRTA allow us to estimate the impact of the non-parabolicity and non-constant relaxation time.

Carrier mobilities were determined using the relation $\mu = \sigma / en$, where $\sigma$ represents the electrical conductivity, $n$ is the carrier density, and $e$ is the electron charge. The calculations were performed with fixed temperature and the chemical potential.
Fig. \ref{fig:g-graphyne-mobilities} shows the corresponding mobilities as a function of carrier concentrations. 
We find SERTA mobilities for electrons and holes in the GY reaching approximately $9.6\times10^3$ ($\approx \times10^4$) and 3.5$\times10^3$ cm$^2/$Vs, respectively. 
The values in the literature for electrons and holes are around 2$\times10^4$ and 4$\times10^3$, respectively~\cite{JXi2014}. 
For GDY, mobility calculations were performed accounting for degenerate bands, yielding approximate values of $7.4\times10^2$ cm$^2$/Vs for electrons and $4.9\times10^2$ cm$^2$/Vs for holes. In comparison, the mobility values reported in the literature are higher, at $\approx 2\times10^5$ cm$^2$/Vs for electrons and $\approx 2\times10^4$ cm$^2$/Vs for holes, respectively~\cite{Gao2011-GDY-mu}. Notably, literature values for electrons are three orders of magnitude higher, while those for holes are two orders of magnitude higher. This discrepancy may be attributed to using the zeroth-order deformation potential theory in Ref.\ \cite{Gao2011-GDY-mu}, which considers only the deformation potentials for a single conduction band and a single valence band in GDY.

\begin{table}[t!]
    \centering
    \caption{\label{table:relaxation-times} Relaxation times $\tau$ in picoseconds (ps).
    The DFT-PBE results for GY and GDY are taken from Ref.~\cite{JXi2014} and Ref.~\cite{Gao2011-GDY-mu}, respectively. The results for GDY represent the average of the relaxation time values computed along two different directions, as reported in Ref.~\cite{Gao2011-GDY-mu} (where GDY-h is 19.11 (15.87) in direction $a$ ($b$), and GDY-e is 1.94 (1.88) in direction $a$ ($b$)). $\tau_{\rm SERTA}$ corresponds to the relaxation time calculated with \textsc{DFTBephy}. $\tau_{\rm CRTA}$, $\tau_{\rm parabolic}$, and $\tau_{\rm Kane}$ represent the effective relaxation-time values. These effective relaxation-time values are determined as the ratio of the maximum SERTA-mobilities to the maximum mobilities obtained within the other approaches, as also indicated in the legend of Fig.\ \ref{fig:g-graphyne-mobilities}.}
    \sisetup{table-format=2.2}
    \begin{tabular}{l S S S S S}
        \toprule
        & \textbf{$\tau_{\rm DFT-PBE}$} & \textbf{$\tau_{\rm SERTA}$} & \textbf{$\tau_{\rm CRTA}$} & \textbf{$\tau_{\rm parabolic}$} & \textbf{$\tau_{\rm Kane}$}  \\
        & \text{[ps]} & \text{[ps]} & \text{[ps]} & \text{[ps]} & \text{[ps]} \\
        \midrule
        \textbf{GY-h}  & 0.52~\cite{JXi2014}  & 0.63  & 0.64  & 0.67  & 0.75  \\
        \textbf{GY-e}  & 1.10~\cite{JXi2014}  & 1.69  & 1.58  & 1.36  & 1.50  \\
        \textbf{GDY-h} & 17.49~\cite{Gao2011-GDY-mu} & 0.04  & 0.08  & 0.07  & 0.08  \\
        \textbf{GDY-e} & 1.91~\cite{Gao2011-GDY-mu}  & 0.14  & 0.10  & 0.10  & 0.11  \\
        \bottomrule
    \end{tabular}
\end{table}

As seen in Fig.\ \ref{fig:g-graphyne-mobilities}, the analytical CRTA results, under the assumption of parabolic bands, exhibit density-independent behavior as expected in 2D. Furthermore, the analytical results from the Kane-band model reasonably align with those from the CRTA and provide a more accurate representation of charge carrier mobility compared to the parabolic-band model. The Kane model shows a good agreement, especially for the holes in GY, consistent with the scattering rates. Hole mobilities in GY remain constant over a broader energy range at lower carrier densities, with scattering rates similarly exhibiting constancy over a large energy range. The deviation between SERTA and Kane-mobilities for electrons in GY becomes apparent at higher carrier densities. A similar trend is also observed for carriers in GDY; however, the deviation arises at lower carrier densities compared to GY. In addition, we calculated the analytical Kane-mobilities using energy-dependent relaxation times $\tau(\varepsilon)$. To have a smooth and continuous representation of the scattering rates as a function of energy, we applied linear interpolation to the scattering rates calculated with \textsc{DFTBephy}. The Kane-mobilities obtained with $\tau(\varepsilon)$ lie between the CRTA and the SERTA results. Thus, incorporating energy-dependent scattering rates provides a more accurate estimation of mobilities in a material, yielding more representative values. This approach enables an analysis of carrier transport properties, offering deeper insights into the interplay between energy-dependent scattering and charge carrier mobility. In Table \ref{table:relaxation-times}, we represent effective relaxation-time values determined as the ratio of the maximum SERTA-mobilities to the maximum mobilities obtained with other models. In the limit of low density, these effective relaxation times values agree with SERTA results.


\section{Conclusions}
\label{sec:coclusions}

We used a computationally efficient DFTB-based approach to calculate electron-phonon couplings of three paradigmatic graphynes and investigated their phonon-limited mobilities.
Benchmarking the DFTB approach against DFT calculations, we studied the structural, electronic and mechanical properties of the $\gamma$-graphynes, GY and GDY, and compared them with graphene. We show that the DFTB approach gives results largely consistent with DFT. Notable differences are found for the band-gaps which are about 1 eV larger than the DFT-PBE values, but are close to calculations with hybrid functionals \cite{GY-GDY-B3LYP,GY-HSE-098,GDY-HSE-083}.

For GY, the relaxation-times calculated within SERTA are 0.63 (1.69) ps for holes (electrons). In the case of GDY, the relaxation times are 0.04 ps and 0.14 ps for holes and electrons, respectively. We find that for holes, scattering rates near the band edge are mainly due to acoustic phonons, while for electrons, optical phonon scattering is the dominant mechanism. Electron scattering rates in GY are approximately $6\times10^{11}$ s$^{-1}$, which is almost the same as intravalley electron scattering rates in graphene  $5\times10^{11}$ s$^{-1}$~\cite{Brandbyge2016-MoS2-Gr}. In contrast, GDY has electron scattering rates of around $7\times10^{12}$ s$^{-1}$, much closer to electron scatterings in MoS$_2$ ($\approx 10^{13}$ s$^{-1}$) \cite{Jacobsen2012-MoS2}.

In the parabolic-band model, analytical mobility calculations based on the CRTA remain independent of carrier concentration. Alternatively, the Kane-band model provides a more accurate description of mobility.
The hole mobilities in GY, calculated with \textsc{DFTBephy}, are on the order of 10$^3$ cm$^2/$Vs, while the electron mobility values reach up to 10$^4$ cm$^2/$Vs. In GDY, the mobility values for both types of charge carriers are on the order of 10$^2$ cm$^2/$Vs.
The phonon-limited mobilities of electrons in GY at room temperature range between those of graphene ($\approx 10^5$ cm$^2/$Vs) \cite{Sarma2008-gr-mobility,Brandbyge2016-MoS2-Gr} and MoS$_2$ ($\approx 400$ cm$^2/$Vs) \cite{Jacobsen2012-MoS2}. On the other hand, the electron mobilities in GDY are of the same order of magnitude as those in MoS$_2$.

The unique properties of GY and GDY, including their tunable porosity, direct band-gaps, and relatively high charge carrier mobilities, render them highly promising for future applications in next-generation nano- and optoelectronics, for energy storage, and for sensing technologies. Further research into their synthesis, stability, and large-scale production will be crucial for exploiting their full potential.


\section{Data availability}
All results we obtained with \textsc{DFTBephy} and \textsc{VASP} packages for graphynes are available in the \textsc{Zenodo} repository, \href{https://zenodo.org/uploads/15113188}{zenodo.org/uploads/15113188}. \textsc{DFTBephy} package is available at \href{https://github.com/CoMeT4MatSci/dftbephy}{github.com/CoMeT4MatSci/dftbephy}.

\section{Declaration of competing interest}
The authors declare that they have no known competing financial interests or personal relationships that could have appeared to influence the work reported in this paper.

\begin{acknowledgments}
We acknowledge the use of computational facilities at the Center for information services and high performance computing (ZIH) at TU Dresden. GC acknowledges funding by the DFG project "Data-Driven Characterization of (A)Chiral 2D Polymers" (CRC1415 - C04).
\end{acknowledgments}

\bibliography{references}

\end{document}



\title{Supplementary Information for Charge Carrier Mobilities in $\gamma$-Graphynes: A computational approach}
\date{\today}

\author{Elif Unsal}
\email{elif.unsal@tu-dresden.de}
\affiliation{Institute for Materials Science and Nanotechnology, TU Dresden, 01062, Dresden, Germany.}

\author{Alessandro Pecchia}
\affiliation{CNR-ISMN, Via Salaria km 29.300, 00015 Monterotondo Stazione, Rome, Italy.}

\author{Alexander Croy}
\affiliation{Institute of Physical Chemistry, Friedrich-Schiller-Universität, 07743, Jena, Germany.}


\author{Gianaurelio Cuniberti}
\email{gianaurelio.cuniberti@tu-dresden.de}
\affiliation{Institute for Materials Science and Nanotechnology, TU Dresden, 01062, Dresden, Germany.}
\affiliation{Dresden Center for Computational Materials Science (DCMS), TU Dresden, 01062, Dresden, Germany.}


\date{\today}

\maketitle
\tableofcontents

\newpage
\section{Structural and Electronic Properties}

\begin{table}[h]
    \centering
    \caption{\label{table-SI-structural-prop.} Calculated structural properties: the lattice constant $a$, the bond lengths (aromatic, single, and triple), and the 2D mechanical moduli ($Y_{2D}$, $B_{2D}$, and $\mu_{2D}$).} 
    \sisetup{table-format=2.2} 
    \resizebox{\textwidth}{!}{  
    \begin{tabular}{l l S S S S S S S}
        \toprule
        & & {$a$ [Å]} & {sp$^2$-sp$^2$ [Å]} & {sp$^2$-sp [Å]} & {sp-sp [Å]} & {$Y_{2D}$ [N/m]} & {$B_{2D}$ [N/m]} & {$\mu_{2D}$ [N/m]} \\
        \midrule
        \textbf{Graphene} & DFTB-mio-1-1 & 2.47 & 1.42 & {-} & {-} & 375 & 250 & 150 \\
                          & DFT-PBE      & 2.47 & 1.43 & {-} & {-} & 328 & 217 & 132 \\
                          & DFT-PBE (Ref.) & {-} & {-} & {-} & {-} & {-} & {-} & {-} \\
        \midrule
        \textbf{GY} & DFTB-mio-1-1 & 6.92 & 1.41 & 1.43 & 1.22 & 174 & 176 & 58 \\
                    & DFT-PBE      & 6.89 & 1.43 & 1.41 & 1.22 & 168 & 141 & 60 \\
                    & DFT-PBE (Ref.~\cite{g-GY-theo-emass}) & 6.89 & 1.43 & 1.41 & 1.22 & 166 & {-} & {-} \\
        \midrule
        \textbf{GDY} & DFTB-mio-1-1 & 9.53 & 1.42 & 1.43 & 1.23 & 141 & 117 & 50 \\
                     & DFT-PBE      & 9.46 & 1.43 & 1.40 & 1.23 & 116 & 99 & 41 \\
                     & DFT-PBE (Ref.~\cite{Gao2011-GDY-mu}) & 9.48 & {-} & {-} & {-} & {-} & {-} & {-} \\
        \bottomrule
    \end{tabular}
    }
\end{table}

\begin{table}[h]
    \centering
    \caption{\label{table-SI-electronic-prop.} Calculated electronic properties: band gap ($E_\text{gap}$) and effective masses ($m_e^*$, $m_h^*$) along $M-\Gamma$ and $M-K$.}
    \sisetup{table-format=1.0}  
    \begin{tabular}{l l S S S}
        \toprule
        & &  & {$M-\Gamma$} & {$M-K$} \\
        & & {$E_\text{gap}$ [eV]} & {$m_e^*$, $m_h^*$ [$m_0$] } & {$m_e^*$, $m_h^*$ [$m_0$] } \\
        \midrule
        \textbf{GY}  & DFTB-mio-1-1 & 1.38 & {0.30, -0.32} & {0.24, -0.26} \\
                     & DFT-PBE      & 0.46 & {0.20, -0.22} & {0.08, -0.09} \\
                     & DFT-PBE (Ref.~\cite{g-GY-theo-emass}) & 0.46 & {0.20, -0.22} & {0.08, -0.09} \\
        \midrule
        \textbf{GDY} & DFTB-mio-1-1 & 1.52 & {0.28, -0.32 } & {0.28, -0.32} \\
                     &              &      & {0.18, -0.20} &  {0.19, -0.21} \\
                     & DFT-PBE      & 0.48 & {0.09, -0.09} & {0.09, -0.09} \\
                     &              &      & {0.07, -0.07} & {0.07, -0.07} \\
                     & DFT-PBE (Ref.~\cite{Gao2011-GDY-mu}) & 0.46 & {-} & {-} \\
        \bottomrule
    \end{tabular}
\end{table}

\newpage

\section{Parametrization Test}

Calculations were performed using the open source DFTB+ code~\cite{DFTB+}. In order to better gauge the effect of the parametrization of the Hamiltonian matrix elements on our results, we used different Slater–Koster sets: 
3ob-3-1~\cite{3ob}
3ob:freq~\cite{3ob-freq}
mio-1-1~\cite{mio-SCC}
ob2-1-1~\cite{ob2}
and matsci-0-3~\cite{matsci}.
We performed a geometry optimization using the SCC method~\cite{mio-SCC}. 

\begin{figure}[h!]
\centering
\includegraphics[width=15.0cm]{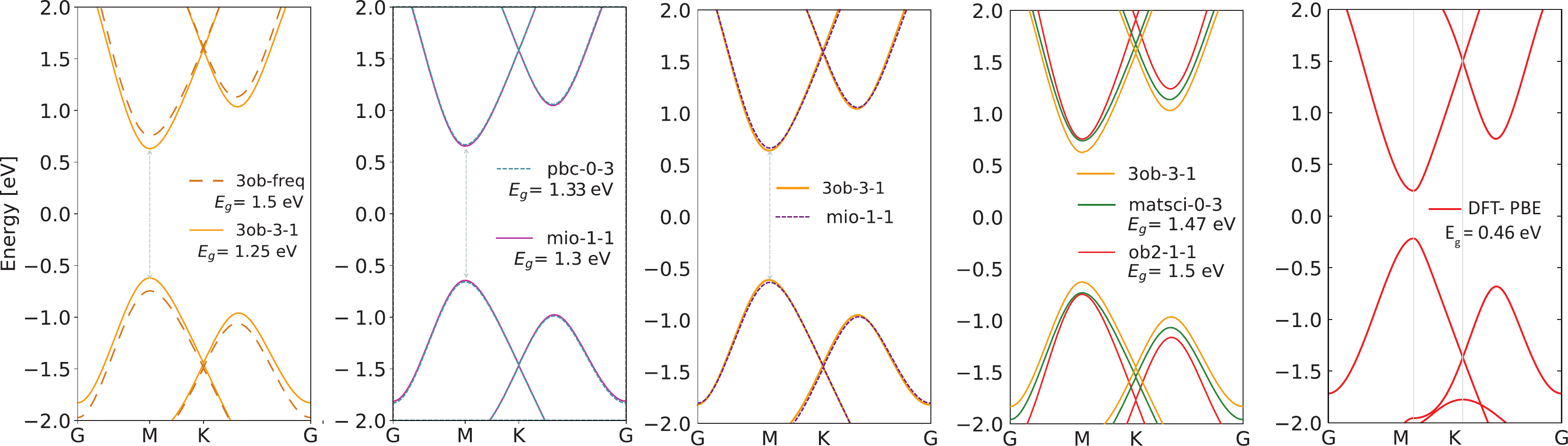}
\caption{Calculated electronic band structures with different Slater-Koster parametrizations.}
\label{fig:SI_e-band}
\end{figure}

\begin{figure}[h!]
\centering
\includegraphics[width=15.0cm]{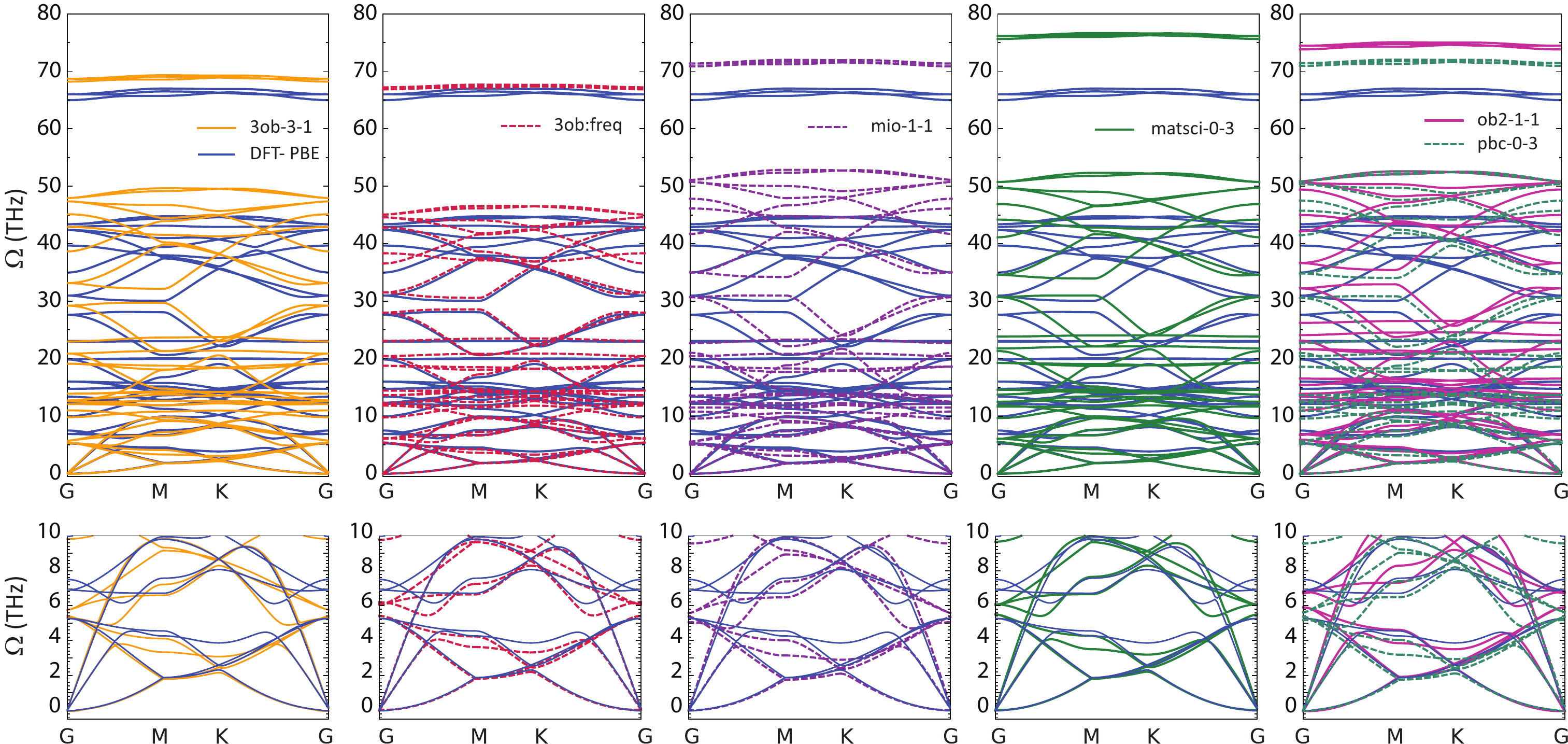}
\caption{Calculated phonon band structures with different Slater-Koster parametrizations.}
\label{fig:SI_p-band}
\end{figure}


\newpage

\section{Kane-Band Model}

For a 2D Kane-band, the electronic dispersion is described by:

\begin{equation}
    \varepsilon + \alpha\varepsilon^2 = \frac{\hbar^2 \mathbf{k}^2}{2m^*},
    \label{eq:Kane-disp}
\end{equation}
where $\alpha$ is the non-parabolicity parameter, $\hbar$ is the reduced Planck constant and $m^*$ is the effective mass of the charge carriers. Here, $\mathbf{k}^2$ is the squared wave vector in 2D, $k_x^2 + k_y^2$.
After rearranging and solving for $\varepsilon$, it can be explicitly written as a function of $\mathbf{k}$

\begin{equation}
    \varepsilon = \left( \frac{\hbar^2 \mathbf{k}^2}{2m\alpha} + \frac{1}{4\alpha^2}\right)^{\frac{1}{2}} - \frac{1}{2\alpha}.  
    \label{eq:Kane-disp2}
\end{equation}


\paragraph{Carrier Velocity}Group velocity of the charge carriers along a certain direction, for instance, in $x$-direction, is defined as the rate of change of the energy $\varepsilon$ with respect to the wave vector $\mathbf{k}$, scaled by the $\hbar$.

\begin{align}
    v_{x} =& \frac{1}{\hbar} \frac{\partial \varepsilon(\mathbf{k})}{\partial k_{x}}  \\
    =& \frac{1}{2\hbar} \left[ \frac{\hbar^2 \mathbf{k}^2}{2m\alpha} + \frac{1}{4\alpha^2} \right]^{-\frac{1}{2}} \frac{2 k_{x} \hbar^2}{2m\alpha} \\
          =& \frac{\hbar k_x}{2m \alpha} \left[ \frac{\hbar^2 \mathbf{k}^2}{2m\alpha} + \frac{1}{4\alpha^2} \right]^{-\frac{1}{2}}\label{eq:velo}
\end{align}

Now, we rearrange the Eq.\ref{eq:Kane-disp2}
\begin{align}
    \left( \frac{\hbar^2 \mathbf{k}^2}{2m\alpha} + \frac{1}{4\alpha^2}\right)^{\frac{1}{2}}= \frac{2\alpha\varepsilon +1}{2\alpha}
    \label{eq:kane-relationv2}
\end{align}
and use this in the Eq.\ref{eq:velo} in order to define velocities in terms of $\alpha$ and energy $\varepsilon$:
\begin{align}
    v_{x} = \frac{\hbar k_x}{2m \alpha}\left( \frac{2\alpha}{2\alpha\varepsilon +1}\right) = \frac{\hbar k_x}{m}\left( \frac{1}{2\alpha\varepsilon +1}\right)
    \label{eq:velo-fin}
\end{align}


\paragraph{Charge Carrier Density}The charge carrier density $n_C$ in 2D at a specific temperature T and chemical potential $\mu$ is given by

\begin{align}
    n_C=& \frac{1}{A_{uc}} \sum _n \int \frac{d^2\mathbf{k}}{\Omega_{BZ}}  \left[ f^0(\varepsilon_n; \mu, T) - f^0(\varepsilon_n; \varepsilon_F, 0)  \right],
\end{align}
where $A_{uc}$ is the unit cell area and $\Omega_{BZ}$ is the Brillouin Zone area, $\Omega_{BZ}=4\pi^2/A_{uc}$. $f^0(\varepsilon;\mu,T)$ is the Fermi distribution function. For the differential area element $d^2\mathbf{k}$  in polar coordinates, we used $2\pi \mathbf{k} d\mathbf{k}$ accounting for the integration over the angular component:

\begin{align}
    n_C=  \frac{1}{4\pi^2} \sum _n \int 2\pi \mathbf{k} d\mathbf{k} \left[ f^0(\varepsilon_n; \mu, T) - f^0(\varepsilon_n; \varepsilon_F, 0)  \right].
\end{align}
where the summation over $n$ accounts for contributions from the bands. The carrier densities for electrons $n_C=n_e$ and holes $n_C=n_h$ are obtained by summing over the conduction and valence bands, respectively. Then we have

\begin{align}
    n_C=  \frac{1}{4\pi^2} \int 2\pi \mathbf{k} d\mathbf{k} \left[ f^0(\varepsilon; \mu, T) - f^0(\varepsilon; \varepsilon_F, 0)  \right].
    \label{eq:nc_polar}
\end{align}

By using the Eq.\ref{eq:Kane-disp}, we can use an expression for $\mathbf{k}$

\begin{equation}
\mathbf{k} = \frac{\sqrt{2m}}{\hbar}\sqrt{\varepsilon (1 + \alpha \varepsilon)},
\end{equation}
and differentiate it with respect to $\varepsilon$ in order to convert the integral from momentum space to energy space.
\begin{align}
    d\mathbf{k} =& \frac{\sqrt{2m}}{\hbar} \left[ \frac{1}{2}(\varepsilon + \alpha\varepsilon^2)^{-\frac{1}{2}} ( 1 + 2\alpha\varepsilon)\right] d\varepsilon \\
                =& \frac{\sqrt{2m}}{2\hbar} \frac{( 1 + 2\alpha\varepsilon)}{\sqrt{(\varepsilon + \alpha\varepsilon^2)}} d\varepsilon               
\end{align}

Next, we substitute the expressions for $\mathbf{k}$ and $d\mathbf{k}$ in to Eq.\ref{eq:nc_polar} to rewrite the integral for carrier densities

\begin{align}
    n_C=&  \frac{1}{2\pi} \int\frac{1+2\alpha\varepsilon}{2\alpha} \frac{2m\alpha}{\hbar^2} d\varepsilon \left[ f^0(\varepsilon; \mu, T) - f^0(\varepsilon; \varepsilon_F, 0)  \right] \\
    =&  \frac{m}{2\pi \hbar^2} \int\  d\varepsilon (1+2\alpha\varepsilon)\left[ f^0(\varepsilon; \mu, T) - f^0(\varepsilon; \varepsilon_F, 0)  \right].
    \label{eq:SI-nC-Kane}
\end{align}


\paragraph{Electrical Conductivity}
Within the Boltzmann transport framework~\cite{Ponce2020}, conductivity tensor in 2D is defined as
\begin{align}
\sigma_{xy} =& \frac{e^2}{A_{uc}} \sum _n \int \frac{d^2\mathbf{k}}{\Omega_{BZ}} \left[ - \frac{\partial f^0(\varepsilon_n; \mu, T)}{\partial \varepsilon}\right] v_{xn} v_{yn} \tau_n,   
\end{align}
where energies $\varepsilon_n$, velocities $v_n$ and relaxation times $\tau_n$ are as a function of $\mathbf{k}$. After summing over the bands, conductivity tensor $\sigma_{xy}$ can be written as
\begin{align}
\sigma_{xy} =& \frac{e^2}{A_{uc}} \int \frac{d^2\mathbf{k}}{\Omega_{BZ}} \left[ - \frac{\partial f^0(\varepsilon; \mu, T)}{\partial \varepsilon}\right] v_{x} v_{y} \tau.   
\end{align}

After inserting $\Omega_{BZ}=4\pi^2/A_{uc}$, $d^2\mathbf{k}= 2\pi \mathbf{k} d\mathbf{k}$ and the velocities as in Eq.\ \ref{eq:velo-fin}, we get
%

\begin{align}
 \sigma_{xy} =&   \frac{e^2}{2\pi} \sum _{x,y} \int \mathbf{k} d\mathbf{k}
\left[ - \frac{\partial f^0(\varepsilon; \mu, T)}{\partial \varepsilon}\right]
\frac{\hbar k_x}{m(1 + 2\alpha \varepsilon)} \frac{\hbar k_y}{m(1 + 2\alpha \varepsilon)}
\tau.
\end{align}

The two-dimensional conductivity $\sigma_{2D}$ is obtained by averaging the trace of the conductivity tensor

\begin{equation}
\sigma_{2D} = \frac{1}{2}tr({\sigma})=\frac{1}{2} \sum_{i=1}^{2}\sigma_{ii} .  
\end{equation}

Since we are only taking the trace, we have $\rightarrow (k_x^2 +k_y^2)/2 = \mathbf{k}^2/2$. We can re-write $\sigma_{2D}$ as

\begin{align}
\sigma_{2D} =&  \frac{e^2}{2 \pi} \int \mathbf{k} d\mathbf{k}
\left[ - \frac{\partial f^0(\varepsilon; \mu, T)}{\partial \varepsilon}\right]
\frac{\hbar^2}{m^{2}(1 + 2\alpha \varepsilon)^2}\frac{\mathbf{k}^2}{2}
\tau. \label{eq:sigma-2D-v1}
\end{align}

Then we change the variable $\mathbf{k}$ to $\varepsilon$ using the following relations.

\begin{align}
\mathbf{k}d\mathbf{k} =& (1 + 2\alpha\varepsilon) \frac{m}{\hbar^2} d\varepsilon \\
\frac{\mathbf{k}^2}{2} =& \frac{m}{\hbar^2}\varepsilon (1 + \alpha \varepsilon)
\end{align}

We substitute these relations in Eq.\ \ref{eq:sigma-2D-v1} and obtain

\begin{align}
 \sigma_{2D} =& \frac{e^2}{2\pi} \int (1 + 2\alpha\varepsilon) \frac{m}{\hbar^2} d\varepsilon 
 \left[ - \frac{\partial f^0(\varepsilon; \mu, T)}{\partial \varepsilon}\right] 
 \frac{\hbar^2}{m^{2}(1 + 2\alpha \varepsilon)^2} \frac{2m}{\hbar^2}(\varepsilon + \alpha \varepsilon^2) \tau.
\end{align}

After simplifying the expression, we arrive at the final form of 2D conductivity within the Kane-band model
\begin{align}
 \sigma_{2D}=& \frac{e^2}{2\pi\hbar^2} \int d\varepsilon 
 \left[ - \frac{\partial f^0(\varepsilon; \mu, T)}{\partial \varepsilon}\right] 
 \frac{(\varepsilon + \alpha \varepsilon^2)}{(1 + 2\alpha \varepsilon)}  \tau.
 \label{eq:SI-sigma}
 \end{align}

The charge carrier densities and electrical conductivity are linearly proportional to the temperature in the parabolic band limit, as $\alpha$ goes to zero. Therefore, the analytical solutions for mobility do not depend explicitly on temperature. When the Kane band approximation is employed, we can see that the charge carrier densities are shifted by $2\alpha\varepsilon$, and the electrical conductivities are no longer linearly related to energy $\varepsilon$, as seen in Eqs.\ \eqref{eq:SI-nC-Kane} and \eqref{eq:SI-sigma}. Therefore, the conductivity and charge carrier density ratio no longer gives a constant and is explicitly temperature-dependent. Therefore, the Kane band model allows for investigating how mobility changes with temperature within the constant relaxation time approximation.

\newpage

\section{Non-Parabolicity in Graphynes}
It is commonly assumed that electronic bands exhibit parabolic dispersion near the band edges. However, deviations from this parabolicity can have a significant impact on the transport properties of materials \cite{Walsh2019,Maassen}. In this study, we employed the quasi-linear Kane-band model\cite{Lundstrom} to investigate nonparabolic effects in graphynes. The non-parabolicity parameter $\alpha$ was determined by fitting the energy dispersion near the band edges. 

\begin{figure}[h!]
\centering
\includegraphics[width=15.0 cm]{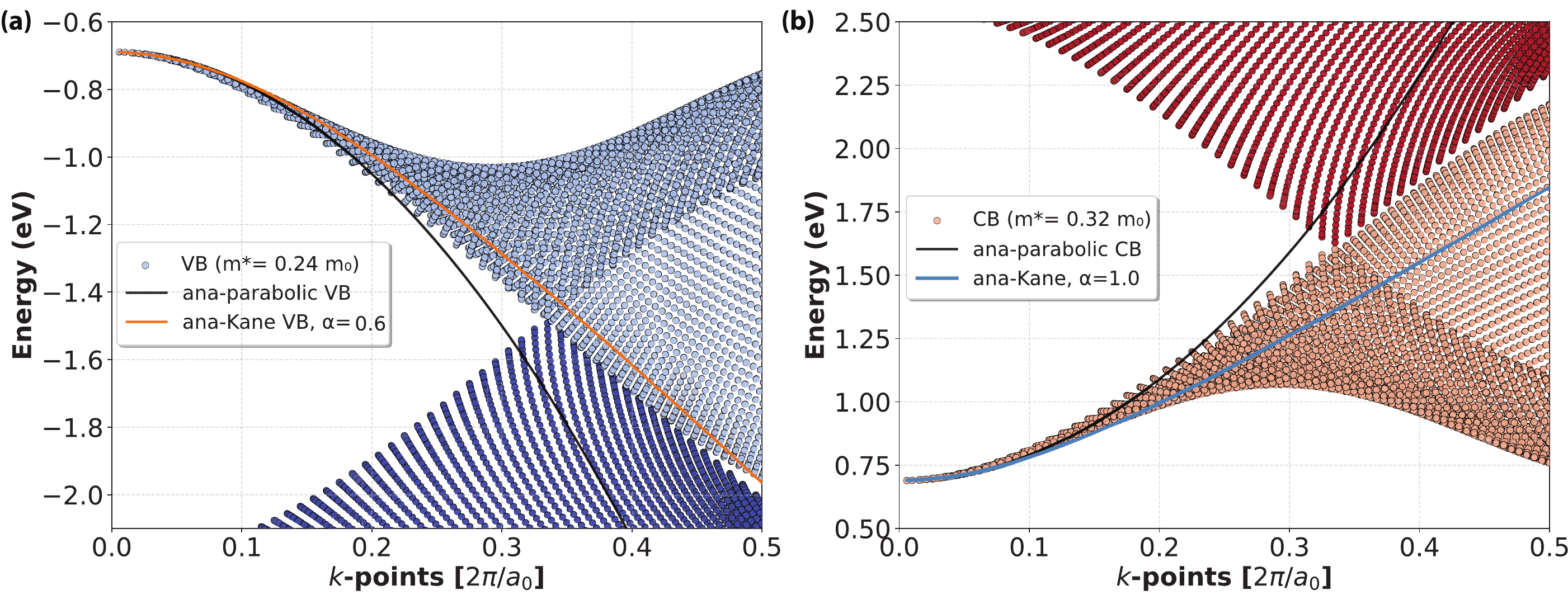}
\caption{(a) Valence and (b) conduction bands graphyne (GY), with the analytical parabolic and Kane bands. }
\label{fig:SI_GY-Kane-fitting}
\end{figure}

\begin{figure}[h!]
\centering
\includegraphics[width=15.0 cm]{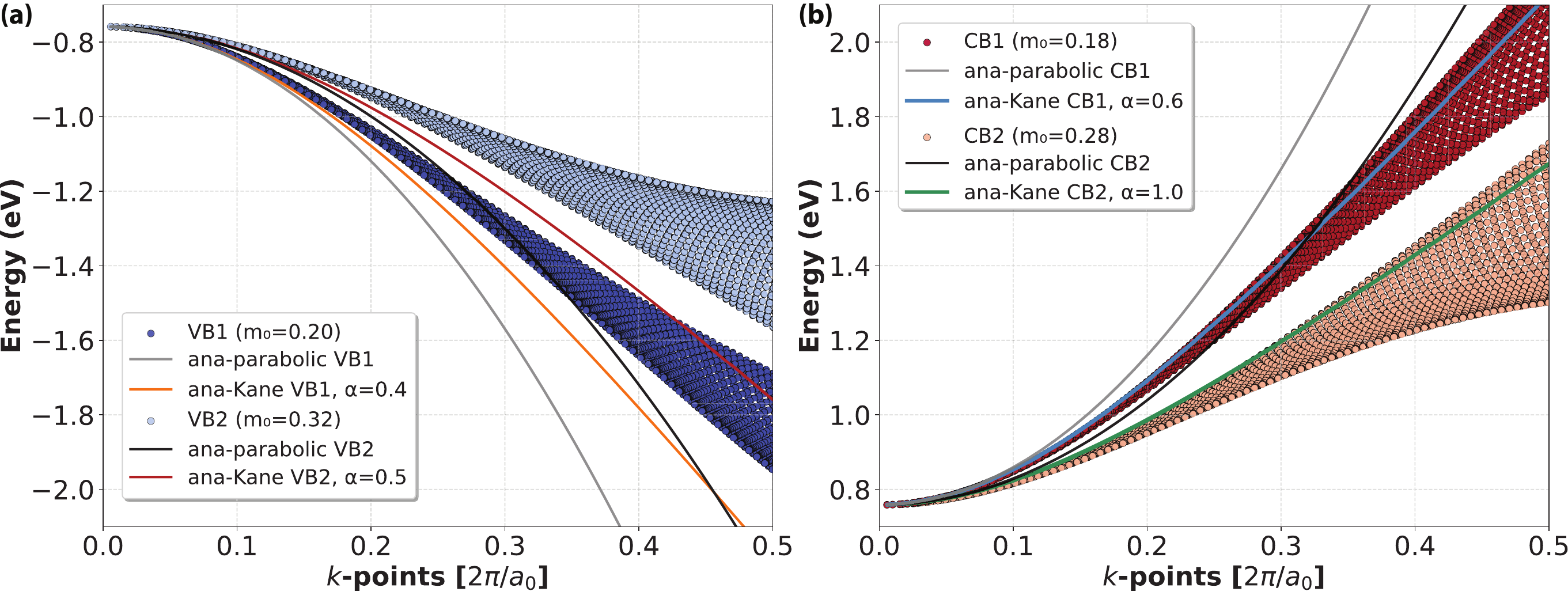}
\caption{(a) Valence and (b) conduction bands in graphdiyne (GDY), with the analytical parabolic and Kane bands.}
\label{fig:SI_GDY-Kane-fitting}
\end{figure}

\newpage

\section{Densities and Electrical Conductivities}

\begin{figure}[h!]
\centering
\includegraphics[width=15.0 cm]{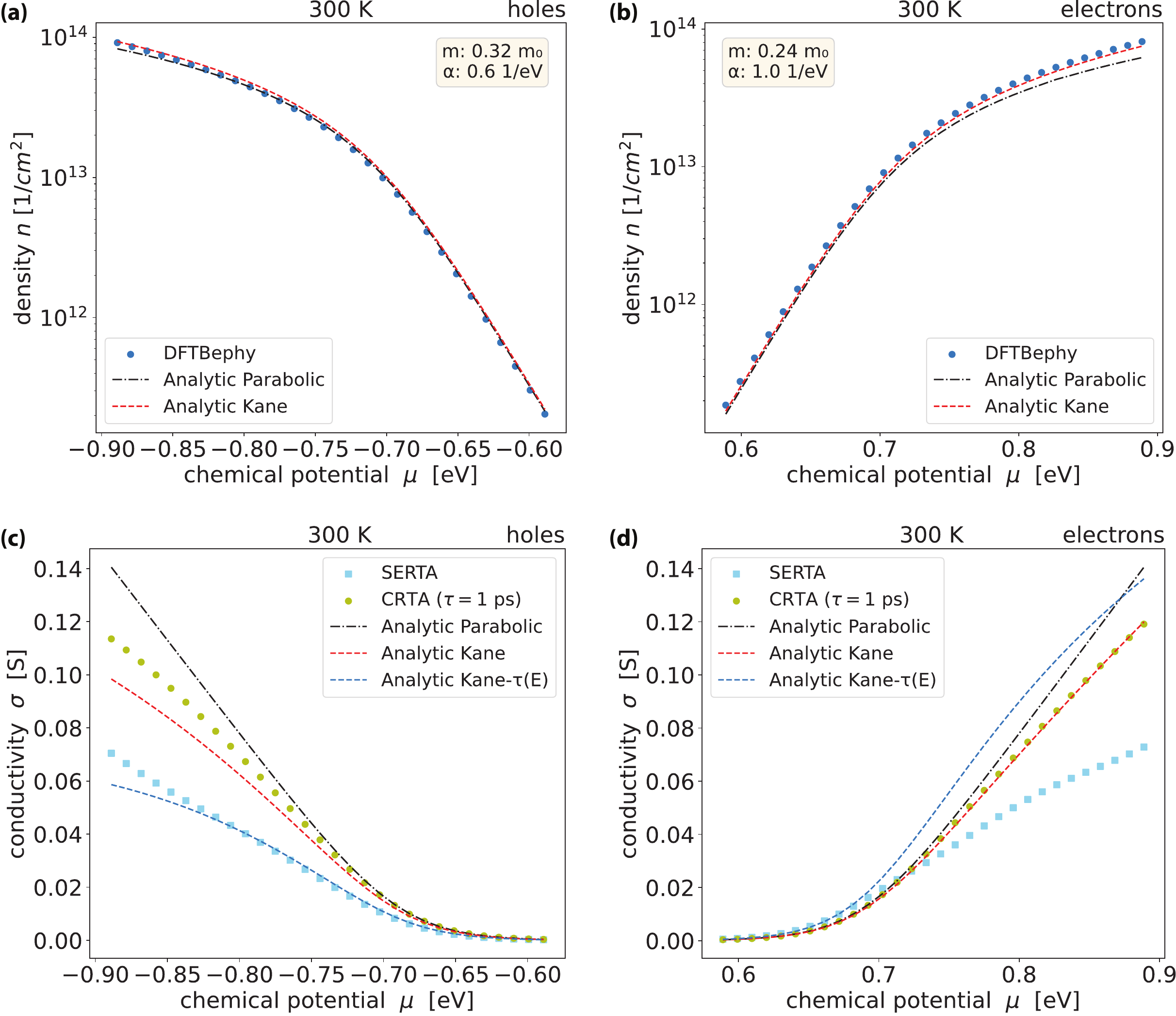}
\caption{Densities for (a) holes and (b) electrons graphyne (GY). Effective mass $m$ and non-parabolicity parameter $\alpha$ values are annotated in the figure. Conductivity values for (c) holes and (d) electrons. }
\label{fig:SI_GY-dos-sigma}
\end{figure}

\begin{figure}[h!]
\centering
\includegraphics[width=15.0 cm]{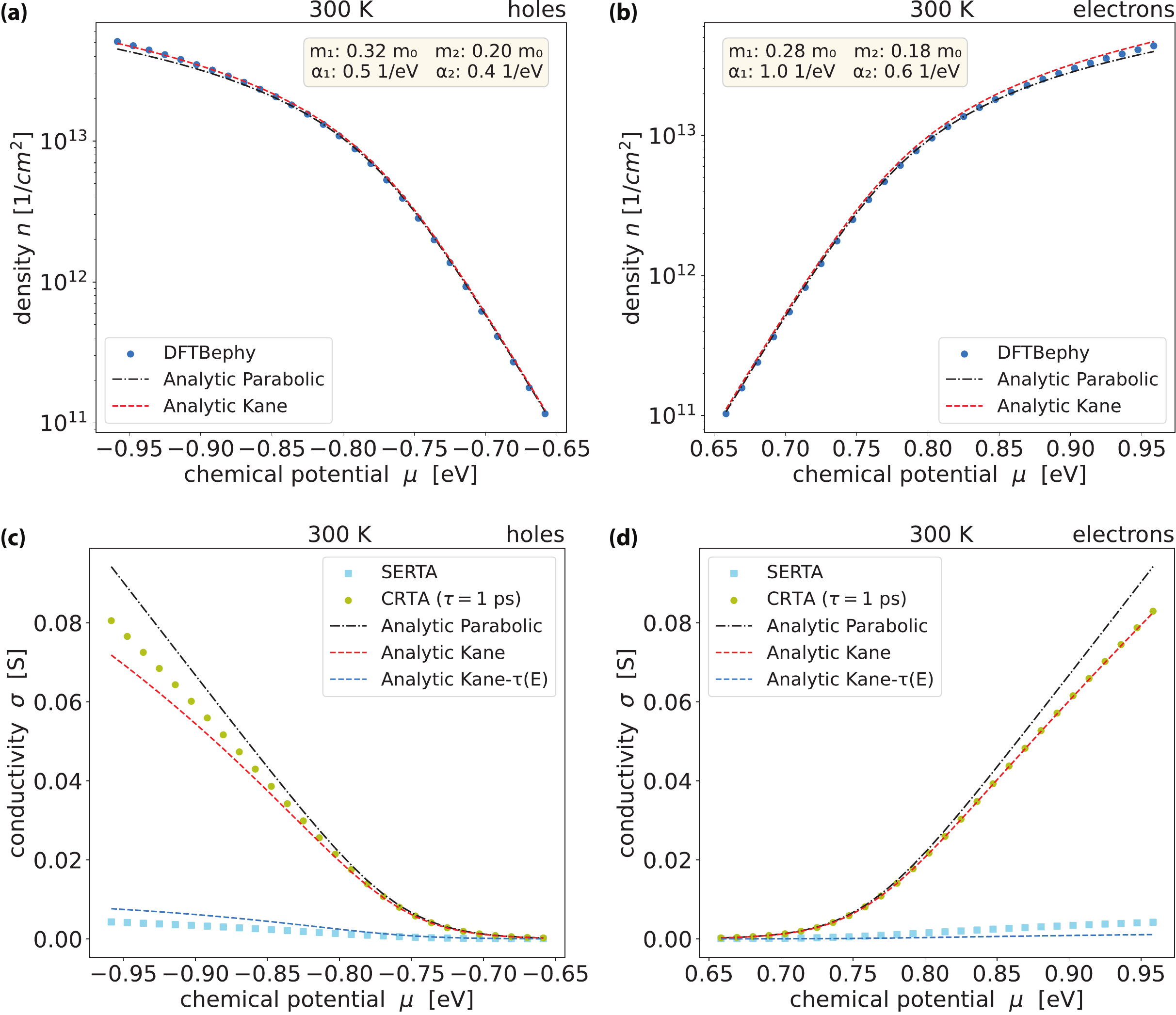}
\caption{Densities for (a) holes and (b) electrons graphdiyne (GDY). Effective mass $m$ and non-parabolicity parameter $\alpha$ values are annotated in the figure. Conductivity values for (c) holes and (d) electrons. }
\label{fig:SI_GDY-dos-sigma}
\end{figure}

The densities and conductivities calculated using DFTBephy, along with the analytical values obtained based on the parabolic and Kane-band models, are presented in Figures \ref{fig:SI_GY-dos-sigma} and \ref{fig:SI_GDY-dos-sigma}. Due to the band-edge degeneracy of the conduction and valence bands in GDY, contributions from these degenerate bands have been taken into account in the density and conductivity calculations. The conductivity values shown in the figures represent the trace $\sigma_x+\sigma_y$.

\newpage

\newpage

\section{Scattering Rates}

\begin{figure}[h!]
\centering
\includegraphics[width=15.0 cm]{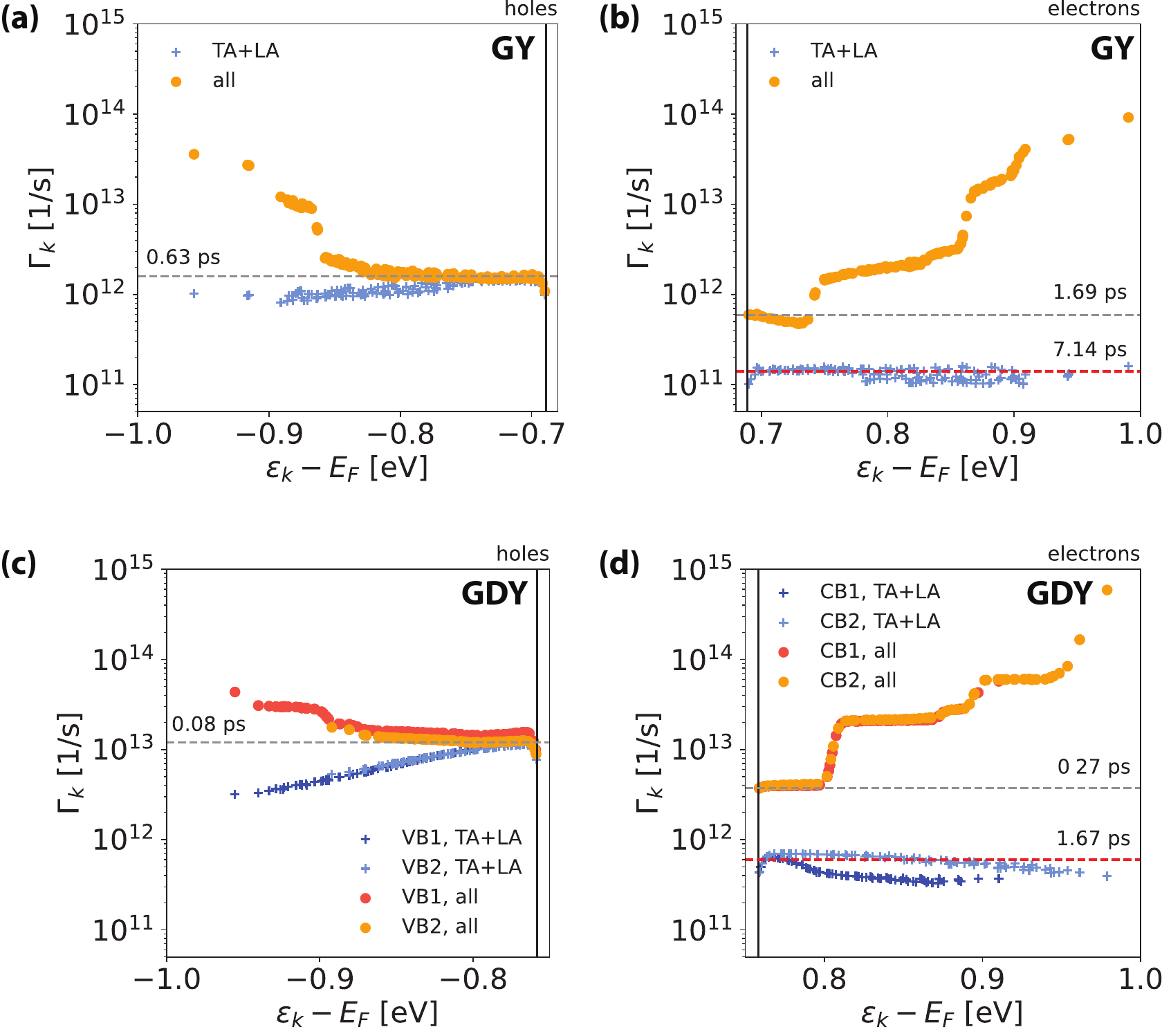}
\caption{The scattering rates (inverse life-times) for (a) the holes and (b) the electrons in monolayer graphyne (GY), as well as for the (c) holes and (d) electrons in monolayer graphdiyne (GDY). In this plot, scatterings were calculated at fixed temperature $T = 300$ K and fixed chemical potential  within SERTA. The chemical potentials were set to the band edge values. }
\label{fig:SI_scatterings}
\end{figure}

\newpage

\bibliography{references}